\documentclass[pdflatex,sn-vancouver-num]{sn-jnl}% Vancouver Numbered Reference Style
\usepackage{graphicx}%
\usepackage{multirow}%
\usepackage{amsmath,amssymb,amsfonts}%
\usepackage{amsthm}%
\usepackage{mathrsfs}%
\usepackage[title]{appendix}%
\usepackage{xcolor}%
\usepackage{textcomp}%
\usepackage{manyfoot}%
\usepackage{booktabs}%
\usepackage{algorithm}%
\usepackage{algorithmicx}%
\usepackage{algpseudocode}%
\usepackage{listings}%

\hypersetup{bookmarksopen=true, bookmarksnumbered=true}
\usepackage{bookmark}
\bookmarksetup{depth=2}

\theoremstyle{thmstyleone}%
\theoremstyle{thmstyletwo}%

\theoremstyle{thmstylethree}%

\begin{document}

\title{Y-Trim: Evidence-gated Adaptase tail trimming for single-stranded bisulfite sequencing}

\author*[1]{\fnm{Yihan} \sur{Fang}}\email{yihan.fang@tufts.edu}

\affil*[1]{\orgdiv{School of Engineering}, \orgname{Tufts University},
\orgaddress{\street{177 College Ave}, \city{Medford}, \postcode{02155}, \state{Massachusetts}, \country{United States}}}

\abstract{
\textbf{Background:} Single-stranded whole-genome bisulfite sequencing (ssWGBS) enables DNA methylation profiling in low-input and highly fragmented material, including cell-free DNA. In widely used post-bisulfite protocols, Adaptase-mediated tailing adds stochastic, template-free end sequence. Unlike adapter-defined junctions, these tails lack a fixed sequence template, so trimming must be decided from FASTQ-stage observables under intrinsic uncertainty.

\textbf{Results:} We show that bisulfite-induced compositional degeneracy implies a \emph{strictly positive} error floor for any fixed per-read boundary rule under a finite nucleotide alphabet. Guided by this limit, we introduce Y-Trim, an evidence-gated framework that separates \emph{admission} (should we trim) from \emph{inference} (where to trim). For Read~2, Y-Trim performs per-read adaptive cut placement via a fixed, chemistry-typed matrix-linear texture scoring scheme; for Read~1, it uses automated sample-level anchoring when read-level localization is feasibility-limited. Across modules, Y-Trim is an explicit, chemistry-specific decision rule with interpretable operating points. On a curated 34-run public cohort (CCGB-34) and simulator stress tests with known latent boundaries, Y-Trim exhibits stable Read~2 operating behavior and Read~1 feasibility-limited behavior consistent with conditional read-through.

\textbf{Conclusions:} Template-free Adaptase tail trimming is best viewed as an evidence-limited FASTQ-stage decision rather than a generic preprocessing knob. By making admissibility and abstention explicit and exposing interpretable genomic-retention versus residual-carryover trade-offs, Y-Trim provides a practical uncertainty-aware preprocessing strategy for ssWGBS.
}




\maketitle

\section{Background}\label{sec:bg}

\begin{figure}[h]
    \centering
    \includegraphics[width=0.96\linewidth]{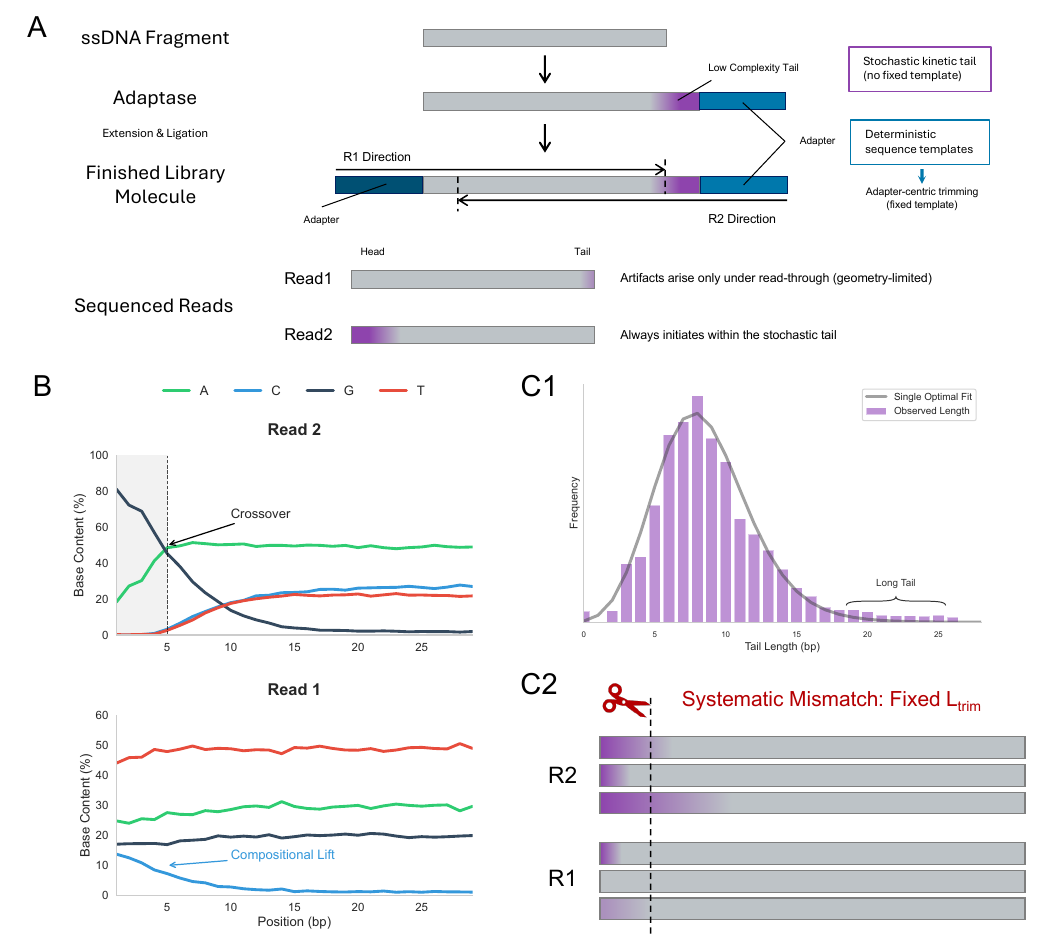}
    \caption{\textbf{Why ssWGBS trimming is an evidence-limited, asymmetric FASTQ-stage decision.}
    \textbf{(A)} Physical origin of post-bisulfite end artifacts and paired-end asymmetry: Adaptase-mediated tailing produces a stochastic low-complexity kinetic tail with no fixed template (purple). This is different from deterministic adapter sequence templates (blue), which are the regime assumed by standard template-based adapter trimming. Read~2 initiates systematically within the tailed region, whereas Read~1 contamination arises only conditionally under insert-length read-through.
    \textbf{(B)} QC-style sample-aggregated per-base sequence content (PBSC) patterns in ssWGBS, shown in an artifact-proximal coordinate. Read~2 shows an early-position composition crossover, while Read~1 exhibits an early-position compositional lift.
    \textbf{(C1)} Read~2 tail lengths are heterogeneous and right-skewed: most reads cluster around a typical range, while a non-negligible fraction extends to much longer lengths. Read~1 is not shown due to conditional insert-length read-through, so there is no direct tail-length variable for it at the FASTQ stage.
    \textbf{(C2)} Consequence of \emph{fixed-length} trimming: any fixed global truncation inevitably mixes over-trimming (loss of genomic bases in short-tail reads) and under-trimming (residual artifacts in long-tail reads).}
    % \Description{Figure 1. ssWGBS trimming is an evidence-limited FASTQ-stage decision: R2 begins in Adaptase tail (systematic) while R1 contamination is conditional read-through; texture evidence includes an indistinguishable regime; R2 tail lengths are long-tailed; fixed cuts over/under-trim.}
    \label{fig:fig1_problem}
\end{figure}

\paragraph{ssWGBS and post-bisulfite end artifacts.}
Single-stranded whole-genome bisulfite sequencing (ssWGBS) enables genome-wide DNA methylation profiling in sample types where conventional double-stranded workflows often fail. ssWGBS is suitable for low-input, highly fragmented, and cell-free DNA sequencing.\cite{lister2009human,luo2021liquid,zeng2018liquid}
However, these advantages come with distinctive preprocessing challenges that arise from post-bisulfite library construction. In widely used post-bisulfite chemistries, Adaptase-mediated tailing introduces synthetic, template-free end sequence whose length and composition are stochastic and chemistry-dependent.\cite{miura2012amplification,miura2019highly,swift_biosciences_tail_trimming_2015,swift_tail_trimming_2019,idt_tail_trimming_2022}
Unlike ligated adapters, which are deterministic and can be removed by standard template-based adapter-trimming logic,\cite{cutadapt,trim_galore,chen2018fastp,bolger2014trimmomatic} Adaptase-driven end artifacts lack a fixed cut-point because their length and composition are heterogeneous across reads.
As a result, the boundary between genomic sequence and synthetic end artifacts can be blurred, and trimming becomes a persistent source of uncertainty across preprocessing and analytical pipelines. Figure~\ref{fig:fig1_problem}A schematizes the physical origin of this post-bisulfite boundary ambiguity.

\paragraph{Paired-end asymmetry.}
A key complication is that post-bisulfite artifacts are \emph{structurally asymmetric} between paired-end reads, driven by insert-length geometry (Fig.~\ref{fig:fig1_problem}A).\cite{swift_biosciences_tail_trimming_2015, swift_tail_trimming_2019, idt_tail_trimming_2022}
Read~2 typically initiates within the Adaptase-tailed region, so synthetic sequence is systematic and boundary handling is unavoidable.
In contrast, Read~1 contains synthetic sequence only conditionally under read-through when read length exceeds insert length (Fig.~\ref{fig:fig1_problem}A, gray), making contamination sparse and geometry-driven rather than ubiquitous (Supplementary Fig.~S1).
This asymmetry implies that a single uniform trimming rule is unlikely to be well-matched to both reads, motivating distinct handling strategies for Read~1 versus Read~2.

\paragraph{FASTQ-level observables: stable sample-level composition signatures.}
Trimming decisions are typically made---and generally must be made---\emph{before alignment}. At this stage, observability is constrained:
beyond the observed base strings (and their aggregate summaries), there is no per-read ground truth for where synthetic sequence ends and genomic sequence begins.
Accordingly, a robust FASTQ-stage summary is the library-level \emph{per-base sequence content} (PBSC) profile aggregated across reads (defined in Methods).\cite{fastqc}
PBSC can exhibit chemistry-consistent signatures even when individual reads remain ambiguous; critically, PBSC is an \emph{aggregate QC signal} and does not provide per-read boundary labels. 

Throughout, we report PBSC in an \emph{artifact-proximal coordinate}: positions are indexed from the putative artifact end.
Concretely, Read~2 is indexed from the read start as sequenced, while Read~1 is indexed from the read end (i.e., using the reversed positional index) so that ``early-position'' refers to the artifact-proximal region for both reads.

Figure~\ref{fig:fig1_problem}B illustrates two canonical PBSC phenomena in post-bisulfite ssWGBS.
For Read~2, PBSC shows an early-position composition crossover consistent with initiation in a low-complexity synthetic regime followed by a transition toward genomic composition.
For Read~1, PBSC is comparatively stable, but a characteristic early-position \emph{compositional lift} can appear when insert-length read-through conditionally exposes end-proximal synthetic sequence.

These geometry- and chemistry-consistent PBSC patterns provide \emph{admissibility screens} for trimming. When signatures are absent, weak, or unstable, FASTQ-stage evidence does not justify trimming at that read end, and we default to non-intervention \emph{without} interpreting this as proof that the library is artifact-free.

% =========================
% Figure 1 (Problem)
% =========================

\paragraph{Long-tailed heterogeneity and the fixed-cut trade-off.}
In real ssWGBS, Read~2 tail lengths are heterogeneous and distinctly right-skewed. While most reads fall in a characteristic range, a non-negligible fraction extends to much longer lengths (Fig.~\ref{fig:fig1_problem}C1). Consequently, no \emph{single} global cut length can be simultaneously appropriate for both short-tail and long-tail reads. Any fixed cut necessarily mixes two structural error modes: \emph{over-trimming} that sacrifices genomic bases in short-tail reads, and \emph{under-trimming} that leaves residual synthetic sequence in long-tail reads (Fig.~\ref{fig:fig1_problem}C2).

We do not plot an analogous Read~1 tail-length distribution, because Read~1 contamination arises only conditionally under insert-length read-through. Therefore, there is no directly comparable per-read \emph{start-of-read} tail-length variable at the FASTQ stage.

\paragraph{Existing ``adaptive'' trimming and the resulting systematic mismatch.}
In practice, common ssWGBS pipelines handle Adaptase-derived sequence by applying a \emph{fixed} truncation. This is because Adaptase tails lack a deterministic sequence template for standard adapter-style matching (Fig.~\ref{fig:fig1_problem}A, blue/template-based regime).\cite{cutadapt,trim_galore,chen2018fastp,bolger2014trimmomatic}
The truncation is often implemented as a small constant (e.g., $\sim$10\,bp) applied uniformly to both Read~2 and Read~1.
Crucially, the constant is typically chosen \emph{a priori} and applied without using FASTQ-stage, sample-level evidence (e.g., PBSC/QC) to assess whether end-artifact signatures are present at a given read end.
When trimming is unnecessary, any fixed cut becomes pure avoidable genomic loss.

This convention is a pragmatic fallback, but it is systematically mismatched to the underlying library-construction chemistry and paired-end sequencing geometry (Fig.~\ref{fig:fig1_problem}A--C). On Read~2, a fixed global truncation therefore commits to an explicit over--under trimming trade-off, sacrificing genomic bases in some reads while leaving residual synthetic sequence in others (Fig.~\ref{fig:fig1_problem}C2). On Read~1, where artifact exposure is conditional on insert-length read-through, applying the same uniform constant further ignores the paired-end non-isomorphism and the fact that contamination is sparse and unsynchronized across reads (Fig.~\ref{fig:fig1_problem}A and C2).
More broadly, a fixed global rule cannot adapt to sample-to-sample heterogeneity in whether and how strongly these chemistry-consistent signatures appear.

\paragraph{Pipeline placement and admissible evidence.}
Trimming choices can affect downstream alignment and methylation quantification.\cite{krueger2011bismark} We therefore study trimming as a \emph{pre-alignment}, FASTQ-stage decision and intentionally avoid post-alignment feedback to prevent circularity: trimming changes alignment outcomes, while alignment-guided trimming presupposes adequate boundary handling. Accordingly, we restrict admissible evidence to chemistry-consistent FASTQ-level summaries (e.g., PBSC). When pre-alignment observables do not admit a defensible boundary placement, we treat abstention or other conservative safeguards as first-class outcomes rather than forcing an irreversible cut.

\paragraph{Overview.}
In this work, we formalize these constraints and introduce an evidence-gated trimming framework that operationalizes them.
We characterize the limits of per-read boundary inference in post-bisulfite ssWGBS, emphasize the fundamental Read~1/Read~2 asymmetry, and propose a conservative approach that acts only when supported by FASTQ-stage evidence.
The resulting operating points explicitly show the trade-off between genomic retention and residual artifact risk imposed by chemistry and paired-end sequencing geometry.

% =========================
% Methods
% =========================
\section{Methods}\label{sec:Methods}

% ---------------------------------------------------------
% 2.1 Scope / pipeline position
% ---------------------------------------------------------
\subsection{Study design and pipeline position (FASTQ-stage, pre-alignment)}
\label{sec:methods:scope}

Y-Trim is specified for paired-end ssWGBS as a FASTQ-stage preprocessing step that runs \emph{before alignment} and \emph{before read merging}.
It operates directly on the raw paired-end FASTQ representation using only the observed base-call alphabet
(\texttt{A/C/G/T/N}) and positional order. \texttt{N} denotes an ambiguous base call in standard FASTQ encoding and is treated as missing where applicable.
No alignment-derived feedback, reference-genome context, or overlap reconstruction is used.

The deployed configuration implements end-specific handling for Read~2 and Read~1 within this pre-merge setting.
Implementation notes for applying the same decision logic on a post-merge representation are provided in Supplementary Note~6.

% ---------------------------------------------------------
% 2.2 Datasets / evaluation
% ---------------------------------------------------------
\subsection{Datasets and evaluation}
\label{sec:methods:data}

\paragraph{CCGB-34 real-data surface.}
We assembled CCGB-34 (Comprehensive Cell-free \& Genomic Bisulfite; $n=34$) as a scenario-indexed ssWGBS evaluation surface spanning
diverse sample contexts and protocol variants.
CCGB-34 is not treated as a read-level ground-truth benchmark; it is curated to expose operational regimes relevant to trimming,
including canonical Read~2 tailing, conditional Read~1 read-through, benign early-position PBSC fluctuations, and negative-control/non-Adaptase
libraries where trimming should abstain.
Scenario definitions and accession manifests are provided in Supplementary Note~8.\cite{G1_li_2024_csf_srp325062,G2_caggiano_2025_als_gse307705,G3_cheng_2021_covid_plasma_srp299418,G4_lo_2025_pancreas_gse249140,G5_cmri_2024_nasal_prjna1162448,G6_heidelberg_2025_heart_prjna1348139,G7_quintanal_2021_lung_prjna1328772}

\paragraph{Chemistry-consistent generative simulator (controlled stress tests).}
Because per-read ground truth is not observable in real ssWGBS without circular alignment-derived proxies, we additionally use a
chemistry-consistent simulator as a controlled stress harness.
The simulator concatenates a post-bisulfite genomic background with a chemistry-typed artifactual prefix governed by a latent contamination-length
variable, enabling robustness checks under known boundaries (Supplementary Note~7.1--7.2). Conceptually, this simulator is a minimal, chemistry-dependent \emph{phenomenological} generative model of template-free end artifacts rather than a mechanical model Adaptase biochemistry.

% ---------------------------------------------------------
% 2.3 Observables + gating
% ---------------------------------------------------------
\subsection{FASTQ-level observables and admission control}
\label{sec:methods:pbsc_gating}

\paragraph{Per-base sequence content (PBSC) and PBSC-style composition statistics.}
We use PBSC as a canonical FASTQ-stage, QC-style summary of position-wise base composition \cite{fastqc}. PBSC-style composition statistics are \emph{aggregated composition summaries} over a certain window. These statistics share the position-wise base-count aggregation with PBSC, and have variations such as non-uniform positional weights. 
Let $p_b(i)$ denote the PBSC at position $i$ for base $b\in\{A,C,G,T\}$:
\begin{equation}
p_b(i) \;=\; \frac{1}{N}\sum_{r=1}^{N}\mathbb{I}\!\left[x_r(i)=b\right],
\end{equation}
where $x_r(i)$ is the base observed at position $i$ of read $r$ (with \texttt{N} treated as missing for PBSC). Positions are interpreted in the artifact-proximal coordinate defined in Background, so that index $i$ refers to the putative artifact-proximal region for both reads. 
PBSC is directly computable from FASTQ prior to alignment and provides a stable library-level position-resolved signature for post-bisulfite
ssWGBS (Fig.~\ref{fig:fig1_problem}B; Supplementary Note~2 and~7.1).

\paragraph{Role of PBSC-style statistics.}
PBSC-style composition statistics are used in three concrete contexts:
(i) as the statistical core of \emph{admission control} (gating) to decide whether trimming analysis is warranted at the sample level (Supplementary Note~2);
(ii) as \emph{reporting coordinates} to summarize position-wise composition before and after trimming in an interpretable, chemistry-consistent way (Supplementary Note~8.2); and
(iii) as inputs to parameterize and populate the chemistry-consistent simulator for controlled stress tests (Methods~\ref{sec:methods:data}, Supplementary Note~7.1).

\paragraph{Admission control (gating).}
Gating operates on PBSC profiles over a fixed inspection window and returns a discrete outcome (\emph{activate} vs.\ \emph{non-activate}). Gating is performed at the sample level. The inspection window is placed in the artifact-proximal region (Read~2 read start; Read~1 read end) under the artifact-proximal coordinate defined in Background. Read~2 admission is based on an early crossover structure in PBSC consistent with a transition from a synthetic regime toward genomic composition. Read~1 admission is formulated as a feasibility check based on a stable early-position compositional lift consistent with conditional read-through
(Fig.~\ref{fig:fig1_problem}B).

An outcome of \emph{non-activate} indicates inadmissibility under FASTQ-stage evidence, not proof that a library is artifact-free.
\emph{Non-activate} is a sample-level gating outcome and should not be conflated with downstream conservative-handling outcomes in the full Y-Trim decision structure (Methods~\ref{sec:methods:ytrim}).
Further clarifications and safeguards are in Supplementary Note~2.

% ---------------------------------------------------------
% 2.4 Y-Trim overview + action modules
% ---------------------------------------------------------
\subsection{Y-Trim decision structure and action modules}
\label{sec:methods:ytrim}

% =========================
% Figure 2 (Solution / Framework)
% =========================
\begin{figure}[t]
    \centering
    \includegraphics[width=1.0\linewidth]{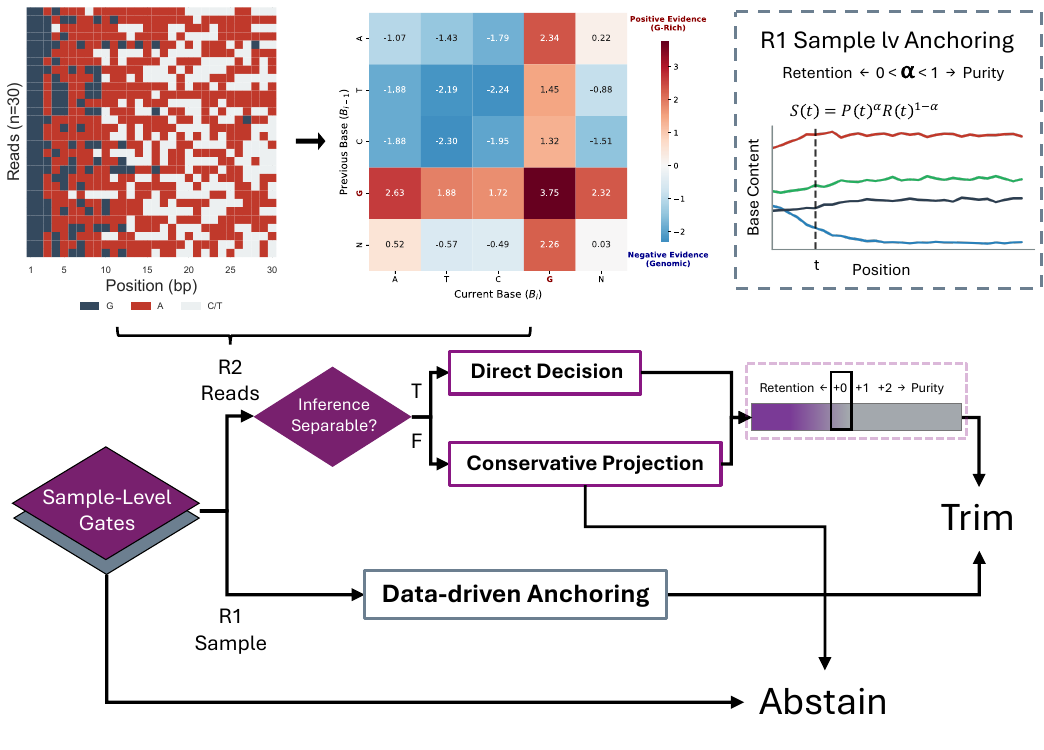}
    \caption{\textbf{Y-Trim: evidence-gated FASTQ-stage trimming with abstention and conservative projection.}
    Sample-level gates decide whether trimming is admissible at each read end from QC-style composition signals.
    If admitted, Read~2 undergoes deterministic texture-based evidence accumulation: separable reads receive a direct cut, while non-separable reads trigger a conservative projection that acts on the same evidence trace without introducing new evidence.
    Read~1 is treated as feasibility-limited and handled by sample-level data-driven anchoring rather than per-read boundary localization.
    Across modules, abstention is a first-class outcome when FASTQ-stage evidence does not justify an irreversible cut.}
    \label{fig:fig2_solution}
\end{figure}

Y-Trim decomposes trimming into (i) admission control (sample-level gating) and (ii) a constrained decision module that is executed only when admission is satisfied (Fig.~\ref{fig:fig2_solution}).
Within (ii), we (a) evaluate evidence under an information-limited FASTQ setting and (b) when a single deployed cutpoint is required, select an operating point that makes the retention--residual risk trade-off explicit rather than implicit.
If admission is not established, Y-Trim abstains from trimming at that read end.

\paragraph{Read~2: texture-based evidence accumulation and stopping.}
After admission, Read~2 is handled at per-read granularity to place an artifact-prefix cut under FASTQ-stage observables (Fig.~\ref{fig:fig2_solution}, R2 module).
Because Adaptase tails have no fixed sequence template, Y-Trim primarily relies on short-range \emph{texture} (low-order local patterns) rather than motif matching.
Under FASTQ-only observables, per-read boundary rules face a strictly positive Bayes error in an indistinguishable regime (Supplementary Note~1), consistent with the non-separability illustrated in Fig.~\ref{fig:fig2_solution} (top left).
Accordingly, Y-Trim is designed for bounded action under admissible evidence rather than ``exact'' boundary recovery.

\paragraph{Shared evidence trace (matrix-linear accumulation).}
For a read with bases $\{b_0,\dots,b_{L-1}\}\subseteq\Sigma$ where $\Sigma=\{A,C,G,T,N\}$, Y-Trim constructs a cumulative evidence trace over prefix length
$p$ within a fixed horizon $L_{\max}$:
\begin{equation}
S(p)
=
\beta_0
+\sum_{i=1}^{p} \mathbf{M}[b_{i-1}, b_i]
+\mathbf{h}^{\top}\mathbf{c}_p
+w_{\text{pos}}\,p,
\qquad 1\le p\le L_{\max}.
\label{eq:methods_r2_evidence_bmc}
\end{equation}
Here $\mathbf{M}$ is a fixed $5\times5$ transition-score matrix over $\Sigma$ encoding short-range texture (Fig.~\ref{fig:fig2_solution}, top),
$\mathbf{c}_p$ is the prefix cumulative base-count vector, and $(\beta_0,\mathbf{h},w_{\text{pos}})$ are fixed stabilizing terms.
The same evidence trace $S(p)$ is used for both direct stopping and conservative projection (Fig.~\ref{fig:fig2_solution}, R2 modules).
All parameters are fixed for a specified chemistry (no per-dataset fitting). Parameterization, design rationale, and intended scope are provided in Supplementary Note~3.

\paragraph{Patience-stabilized crossing and conservative projection.}
A direct cut is produced only when $S(p)$ indicates sustained departure from the synthetic-dominated regime.
With patience parameter $\pi$, Y-Trim returns the earliest prefix length at which $S(p)$ remains non-negative for $\pi+1$ consecutive positions
(Supplementary Note~3).
All Read~2 inference is constrained to a hard horizon $L_{\max}=25$ bases (epistemic bound, not an estimate of true tail length; Supplementary Note~3).
If no stable stopping decision occurs within $L_{\max}$, the read is treated as non-separable under the fixed horizon and handled by conservative projection that acts only on the already-computed trajectory $S(p)$ and introduces no new evidence (Supplementary Note~4).

\paragraph{Read~1: sample-level anchoring (feasibility-limited setting).}
Read~1 artifact exposure is geometry-driven and conditional: insert-length read-through acts like a latent truncation/shift on the observed sequence, so any end-artifact signal is sparse and unsynchronized across reads rather than forming a shared local transition.
As a result, Read~1 evidence is typically unstable at the per-read level and becomes interpretable primarily under aggregation, making per-read boundary localization frequently ill-posed.
Accordingly, Y-Trim handles Read~1 via sample-level anchoring that returns a single trimming constant per sample when Read~1 intervention is admitted
(Fig.~\ref{fig:fig2_solution}, R1 module; Supplementary Note~5).
Candidate trim lengths are evaluated within a bounded search space ($\le 25$ bases), and the deployed configuration uses a fixed purity--retention trade-off
weight ($\alpha=0.5$) without per-sample tuning; the constant is inferred from FASTQ-stage evidence under this fixed objective (Supplementary Note~5).

\paragraph{Deployment semantics and operating points.}
Y-Trim is a chemistry-typed, white-box decision rule with a small number of interpretable operating-point parameters.
For Read~2, the deployment offset selects among operating points along a common decision geometry without changing the underlying evidence evaluation (Figure~\ref{fig:fig2_solution}, R2 module).
For Read~1, $\alpha$ controls the meaning of the sample-level anchoring objective and is therefore treated as part of the method definition rather than a routine tuning knob (Figure~\ref{fig:fig2_solution}, top right).
Further discussion of operating semantics and their interpretation is provided in Supplementary Notes~6 and~9.

% ---------------------------------------------------------
% 2.5 Baselines + reporting + software
% ---------------------------------------------------------
\subsection{Baselines, reporting summaries, and software}
\label{sec:methods:baselines_reporting_software}

\paragraph{Baselines and reference variants.}
On CCGB-34, we report three classes of preprocessing behavior: (i) the deployed Y-Trim configuration (our primary method, Figure~\ref{fig:fig2_solution}),
(ii) a practical fixed-length baseline, and (iii) explanatory development-time reference variants
(Supplementary Notes~3--6).
For (ii), we use symmetric 10\,bp fixed trimming on both reads as a practice-driven reference point. This is a commonly documented and deployed configuration rather than as a ground-truth target.
For (iii), we include a small set of reference variants used only for ablation-style interpretation (not for competition): 
a Read~2 anchoring-style reference variant, a Read~2 stopping/projection reference variant family, and Read~1 per-read ``adaptive'' attempts contrasted against the deployed sample-level anchoring. 
All methods and reference variants are evaluated on CCGB-34 under the same reporting summaries below.

\paragraph{Reporting summaries (over- vs.\ under-trimming directions).}
For real-data comparisons on CCGB-34, we summarize trimming behavior in a fixed two-axis coordinate system that separates the two structural error modes of
boundary decisions: (i) avoidable genomic loss due to over-trimming and (ii) residual artifact carryover due to under-trimming.
Axis definitions and implementation are provided in Supplementary Note~8.
We additionally report an end-proximal excess-signal aggregation to connect trimming to a methylation-relevant endpoint without requiring biological labels
(Supplementary Note~7).

\paragraph{Software and reproducibility.}
Y-Trim is provided as a transparent Python reference implementation together with scripts to reproduce all analyses and figures.
For a specified chemistry, all parameters are fixed at deployment (no training or dataset-adaptive estimation), leaving runtime behavior deterministic and
reproducible.
Implementation details, I/O semantics, and supported scope are documented in Supplementary Note~6.

% =========================
% Results
% =========================
\section{Results}\label{sec:results}

% =========================
% Figure 3 (Robustness + Frontier + CpG stress)
% =========================
\begin{figure}[t!]
    \centering
    \includegraphics[width=0.85\linewidth]{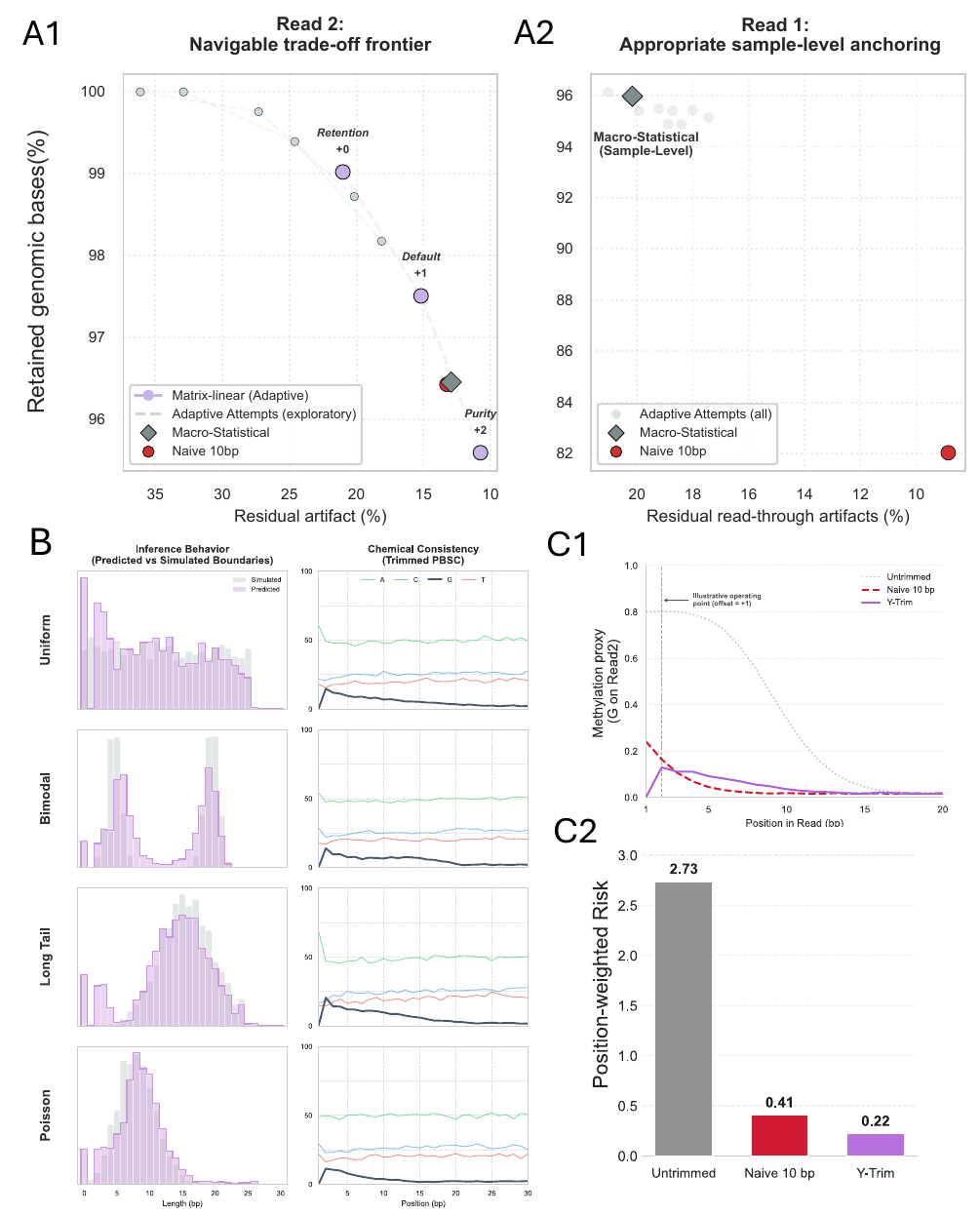}
    \caption{\textbf{Asymmetric decision geometry and stability under regime shift with end-proximal, CpG-sensitive stress.}
    \textbf{(A1)} On admitted CCGB-34 Read~2 samples, evidence-gated trimming reveals a clear retention--residual trade-off geometry; the deployment offset selects interpretable operating points along this fixed decision interface.
    \textbf{(A2)} On admitted Read~1 samples, read-level ``adaptive'' cut placement does not form a coherent frontier and instead collapses into an unstructured cluster, consistent with geometry-limited feasibility; this motivates conservative sample-level anchoring and sample-level abstention.
    \textbf{(B)} Simulator-backed out-of-distribution stress tests: across diverse contamination-length families and fixed decision parameters, Y-Trim exhibits stable, ordered behavior rather than collapsing to a single default cut.
    \textbf{(C)} CpG-motivated stress evaluation emphasizes end-proximal sensitivity in methylation contexts:
    \textbf{(C1)} post-trim Read~2 dominant-\emph{G} signal is suppressed in a chemistry-consistent manner, and
    \textbf{(C2)} an end-proximal, position-weighted excess-signal score compares Y-Trim against a commonly used fixed-length reference (10\,bp) under CpG-sensitive risk aggregation.}
    \label{fig:fig3_robust}
\end{figure}

\subsection{CCGB-34 reveals asymmetric feasibility: Read~2 admits frontier navigation, Read~1 does not}
\label{sec:results:ccgb_asym}

We evaluated Y-Trim on CCGB-34 using common reporting axes (Supplementary Note~8), summarizing only samples admitted at the corresponding read end.
This separates deployment behavior from cases where FASTQ-stage evidence does not justify intervention. For the CCGB-34 analyses shown in Fig.~\ref{fig:fig3_robust}A--C, the admitted subsets contain 15 samples for Read~2 and 15 samples for Read~1 under the corresponding end-specific gates; exact scenario membership, accession manifests, and admission outcomes are listed in Supplementary Note~8. Replication scripts and intermediate summaries are included in the accompanying code package.

\paragraph{Terminology (plot labels).}
In Fig.~\ref{fig:fig3_robust}, \emph{Matrix-linear} denotes the Read~2 per-read module that applies the shared evidence trace and stopping/projection logic, while \emph{Macro-statistical} denotes the sample-level anchoring module that operates under aggregation when per-read localization is not feasible.
Both are components of a single method (Y-Trim) and share the same evidence-gated interface; they are shown as separate labels only to reflect the distinct feasible action granularity on Read~2 versus Read~1.

\paragraph{Read~2 forms an interpretable operating frontier.}
On admitted Read~2 samples, trimming yields a coherent trade-off between residual synthetic signal and retained genomic bases (Fig.~\ref{fig:fig3_robust}A1).
This frontier reflects the evidence-limited nature of FASTQ-stage boundary decisions: under intrinsic non-separability, no single FASTQ-stage decision rule can simultaneously eliminate all residual artifact while preserving all genomic bases (Supplementary Note~1).
Varying the deployment offset traverses this frontier without changing the underlying evidence evaluation, producing a small set of interpretable operating points.
A fixed 10\,bp trim locates in a plausible region of the same geometry, consistent with its role as a pragmatic default, but it lacks admission semantics and cannot accommodate long-tailed heterogeneity.
Details on projection behavior and its operational interpretation are provided in Supplementary Note~9.

\paragraph{Read~1 is feasibility-limited at read level; naive fixed trimming can be damaging.}
In contrast, Read~1 does not yield a meaningful read-level frontier (Fig.~\ref{fig:fig3_robust}A2).
Read-level ``adaptive'' attempts concentrate into an unstructured cluster, consistent with conditional read-through and lack of per-read synchronization.
In this regime, a uniform fixed cut for both reads can become a high-cost intervention: the 10\,bp reference achieves low residual artifact only by incurring substantial avoidable genomic loss, illustrating why treating Read~1 as ``a weaker Read~2'' is structurally mismatched in post-bisulfite libraries.
Y-Trim therefore defaults to conservative sample-level anchoring and abstention for Read~1 rather than forcing per-read boundary navigation.

\subsection{Simulator-backed OOD stress tests show stable behavior under regime shift}
\label{sec:results:sim}

To probe behavior under controlled regime shifts, we used a minimal chemistry-consistent simulator with known latent boundaries (Supplementary Note~7.1).
Across out-of-distribution contamination-length families, Y-Trim remains stable under fixed parameters (Fig.~\ref{fig:fig3_robust}B):
outputs shift in an ordered, structured manner with the imposed length regimes rather than collapsing to a single default cut or drifting unpredictably.
This is consistent with robustness of the deployed FASTQ-stage decision interface to tail-length regime changes.

\subsection{End-proximal CpG-sensitive aggregation highlights reduced methylation-relevant risk}
\label{sec:results:risk}

In methylation contexts, small end-proximal artifacts can have disproportionate impact, motivating a risk-emphasis view of trimming outcomes.\cite{deaton2011cpg,siegfried2010dna}
We therefore summarize post-trim behavior using Read~2 dominant-\emph{G} suppression and an end-proximal, position-weighted excess-signal score (Supplementary Note~7.3).

\paragraph{Content-aware suppression versus a fixed translation.}
Fig.~\ref{fig:fig3_robust}C1 shows that a fixed-length trim acts as a content-agnostic translation of the profile, which can leave residual artifact beyond the cut under heterogeneity.
In contrast, Y-Trim suppresses dominant-\emph{G} signal in a chemistry-consistent manner, targeting the end-proximal region most likely to drive downstream bias.

\paragraph{CpG-sensitive risk aggregation makes operating-point trade-offs explicit.}
We summarize end-proximal behavior using a position-weighted excess-signal score (Supplementary Note~7.3).
Under this aggregation, Y-Trim improves upon the untrimmed baseline and is competitive with a commonly used 10\,bp reference, while making the purity--retention trade-off explicit through its operating points (Fig.~\ref{fig:fig3_robust}C2).
We treat the fixed-length reference as a practice-driven coordinate, not as a uniquely correct FASTQ-stage quantity.

\section{Discussion}\label{sec:discussion}

\paragraph{Trimming is an evidence-limited FASTQ-stage decision.}
Trimming in post-bisulfite ssWGBS is often treated as a generic preprocessing knob, but our results emphasize that it is fundamentally a \emph{FASTQ-stage} decision problem: the only universally admissible evidence is the observed base string and
its position.
Under bisulfite-induced compositional collapse, FASTQ-only observables can enter an intrinsically indistinguishable regime in which genomic and synthetic sequence cannot be separated from a finite nucleotide alphabet, implying a \emph{strictly positive} error floor for any fixed per-read boundary rule (Supplementary Note~1).
This shifts the appropriate objective away from ``exact'' boundary recovery and toward \emph{bounded, chemistry-consistent risk control} under explicit uncertainty.\cite{siegfried2010dna}

\paragraph{Asymmetry creates a frontier in Read~2 but not in Read~1.}
Taking the data-generating asymmetry seriously explains the qualitative structure seen in CCGB-34. Read~2 carries systematic Adaptase-derived signal at the read start, so admitted samples exhibit a coherent, navigable decision geometry: a small deployment offset traverses an interpretable frontier between residual end-artifact carryover and genomic-base retention \emph{without changing} evidence evaluation (Fig.~\ref{fig:fig3_robust}A1; Supplementary Notes~2--3). Read~1 behaves differently: contamination is conditional on insert-length read-through and becomes coherent primarily under aggregation, so read-level adaptivity collapses into an unstructured cluster rather than a frontier (Fig.~\ref{fig:fig3_robust}A2). This is not a weaker version of Read~2, but a different inferential regime; treating Read~1 as ``weaker R2'' is precisely what produces the pathological behavior of symmetric fixed-length trimming on Read~1.

\paragraph{Feasibility before inference: gate, bounded Read~2 action, anchored Read~1.}
These constraints motivate a staged design in which \emph{feasibility is assessed before inference} and irreversible action is taken only under admissible evidence. In Y-Trim, sample-level admission (gating) serves as a non-intervention principle, preventing trimming when PBSC structure is weak, inconsistent, or plausibly explained only by aggregation effects (Supplementary Note~2). After admission, Read~2 uses a fixed-rule (non-learning) evidence accumulation with a patience-stabilized stopping requirement and an explicit hard horizon to avoid false precision once discriminative information becomes non-separating (Supplementary Note~3). When direct placement is not supported by admissible evidence, uncertainty is handled as a first-class outcome via conservative projection rather than by escalating aggressiveness (Supplementary Note~4). Read~1 is handled by sample-level anchoring that resolves a purity--retention trade-off under geometry-limited uncertainty without claiming per-read boundary localization (Supplementary Note~5).

\paragraph{Deployment and transfer.}
A practical implication is that deployment should make the remaining degrees of freedom explicit. Offset is a recommended operating-point selector that moves along an existing Read~2 decision geometry \emph{without} changing evidence evaluation, whereas $\alpha$ changes the meaning of the Read~1 anchoring objective itself and is therefore
not a routine tuning knob (Supplementary Note~6).
More broadly, the evidence-gated framing is not specific to ssWGBS: it applies \emph{when} end artifacts are template-free but texture-defined,
and the synthetic process shares the same finite observable alphabet as biological sequence. In such settings, transfer does not require a shared motif but rather a chemistry-consistent, locally observable statistical signature and a
validation path (e.g., simulator-backed stress) that prevents over-claiming in information-limited regimes (Supplementary Note~7).

\section{Conclusion}\label{sec:conclusion}

Single-stranded bisulfite sequencing enables methylation profiling in low-input and fragmented samples, but post-bisulfite library construction introduces stochastic, template-free end artifacts that limit what can be inferred about read boundaries
from FASTQ observables alone.Because FASTQ-only observables can enter an intrinsically indistinguishable regime under a finite nucleotide alphabet, there is a strictly positive error floor for any fixed per-read boundary rule (Supplementary Note~1). This structure explains both why fixed-length trimming persists as a pragmatic default and why it is systematically mismatched to heterogeneous, right-skewed tailing behavior.

Y-Trim operationalizes an evidence-gated alternative: it separates \emph{admissibility} from \emph{cut placement}, treats Read~2 and Read~1 according to their distinct physical origins, and exposes interpretable operating points for conservative deployment. Using CCGB-34 as a real-data evaluation surface and a chemistry-consistent simulator for controlled stress tests, we show that Read~2 admits a navigable genomic-retention versus residual-carryover trade-off geometry, while Read~1 is feasibility-limited and is better
handled via sample-level anchoring and abstention. 
Together, these results support uncertainty-aware preprocessing for ssWGBS that makes admissibility, operating points, and conservative handling explicit rather than implicit.

\backmatter

\vspace{2cm}

\section*{Declarations}

\subsection*{Ethics approval and consent to participate}
Not applicable.

\subsection*{Consent for publication}
Not applicable.

\subsection*{Availability of data and materials}
No new sequencing data were generated for this study.
All sequencing datasets analyzed are publicly available from NCBI repositories (GEO/SRA/BioProject).
The CCGB-34 cohort (Comprehensive Cell-free \& Genomic Bisulfite; $n=34$) was assembled from seven published studies and associated accessions:
GSE178666/SRP325062 \cite{G1_li_2024_csf_srp325062},
GSE307705/SRP619043 \cite{G2_caggiano_2025_als_gse307705},
SRP299418 \cite{G3_cheng_2021_covid_plasma_srp299418},
GSE249140/SRP475142 \cite{G4_lo_2025_pancreas_gse249140},
PRJNA1162448/SRP533334 \cite{G5_cmri_2024_nasal_prjna1162448},
PRJNA1348139/SRP636882 \cite{G6_heidelberg_2025_heart_prjna1348139},
and PRJNA1328772/SRP620537 \cite{G7_quintanal_2021_lung_prjna1328772}.
Supplementary Notes, figures, and tables are available with this article (in the submission/review package).

A complete reference implementation of Y-Trim (core inference engine, simulator, demo, and full figure-reproduction workflows) is available for peer review via a view-only OSF link: \url{https://osf.io/etqrj/overview?view_only=795c81dbfa8a4f7185b1c45310628591}. The OSF package includes scripts to reproduce all analyses and figures reported in the manuscript and includes selected intermediate artifacts necessary for reproduction. 

A preprint version of this work is available on arXiv: \url{https://doi.org/10.48550/arXiv.2601.19002}.
The submitted manuscript revises the narrative organization and updates figures for improved coherence with the target readership and journal scope.

\subsection*{Competing interests}
The authors declare no competing interests.

\subsection*{Funding}
This work was conducted without external grant funding.
Publication-related costs are supported through institutional open access resources.

\subsection*{Authors' contributions}
Y.F.\ conceived the project, developed the theoretical framework and algorithms, performed all computational analyses, and wrote the manuscript.

\subsection*{Acknowledgements}
We are deeply grateful to Prof. Lenore Cowen for sustained mentorship, critical guidance on research direction and positioning, and extensive feedback throughout the development of this work. We are grateful to Dr. Yongkun Ji for discussions that clarified the ssWGBS chemistry and pipeline context, and for feedback that improved domain framing and technical conventions. We thank Dr. Rebecca Batorsky for detailed feedback on validation strategy and presentation, and for emphasizing the importance of evaluating downstream biological consequences during method development. We thank Jindan Huang for valuable advice on research planning and publication strategy. We thank Dr. William White for helpful discussions, broader scientific feedback, and advice on manuscript presentation. We thank Prof. Jivko Sinapov for early developmental discussions and algorithmic perspective. We also thank Dr. Albert Tai for technical discussions regarding sequencing workflows, as well as Hiu Mai and Yijia Zhang for preliminary literature exploration.

During preparation of this manuscript, the author used generative AI tools for language editing and for assistance with software development tasks (e.g., testing and debugging analysis scripts). All AI-suggested text and code were reviewed, edited, and validated by the author, who takes full responsibility for the content of the manuscript and the integrity of the analyses.

\bibliography{sn-bibliography}

\end{document}